\documentstyle[aas_macros]{PoS}

\title{Gamma-ray pulsars with Fermi}

\ShortTitle{Gamma-ray pulsars with Fermi}

\author{\speaker{David A. Smith}%
        \thanks{for the Fermi LAT collaboration and Pulsar Timing Consortium}\\
       Centre d'\'Etudes Nucl\'eaires de Bordeaux Gradignan, IN2P3/CNRS, Universit\'e Bordeaux 1, BP120, F-33175 Gradignan Cedex, France\\
       E-mail: \email{smith@cenbg.in2p3.fr}}

\author{Lucas Guillemot\\ Laboratoire de Physique et Chimie de l'Environnement et de l'Espace -- Universit\'e d'Orl\'eans / CNRS, F-45071 Orl\'eans Cedex 02, France and Station de radioastronomie de Nan\c{c}ay, Observatoire de Paris, CNRS/INSU, F-18330 Nan\c{c}ay, France}
\author{Matthew Kerr\\ Space Science Division, Naval Research Laboratory, Washington, DC 20375-5352, USA\\
CSIRO Astronomy and Space Science, Australia Telescope National Facility, Epping NSW 1710, Australia
}
\author{Cherry Ng\\   Department of Physics and Astronomy, The University of British Columbia, Vancouver, BC V6T-1Z1, Canada}
\author{Ewan Barr\\   Max-Planck-Institut f\"{u}r Radioastronomie, Auf dem H\"{u}gel 69, 53121 Bonn, Germany}

\abstract{In 8 years of operation, the Large Area Telescope (LAT) on the {\em Fermi} satellite has impacted our understanding of gamma-ray pulsars dramatically. 
The LAT now sees over two hundred pulsars: the largest class of GeV sources in the Milky Way. 
They are diverse -- radio loud versus quiet, young versus millisecond, in evolving binary systems versus isolated, and so on. 
Relatively few of the GeV pulsars have also been seen in soft gamma rays.
After an overview, we present 10 new radio pulsars, six young and four recycled, for which we detect gamma-ray pulsations.}

\FullConference{11th INTEGRAL Conference Gamma-Ray Astrophysics in Multi-Wavelength Perspective\\
         10-14 October 2016\\
         Amsterdam, The Netherlands}

\begin{document}

\section{The Large Area Telescope on {\em Fermi}}
On 11 June, 2008 a Delta-II rocket blasted off from Cape Canaveral and placed the GLAST satellite in low orbit. 
All systems performed nominally (as they still do) and GLAST was renamed to honor Enrico Fermi. 

{\em Fermi}'s primary instrument is the Large Area Telescope, or LAT. 
Its field of view covers 20\% of the sky, with a large collection surface of roughly (150 cm)$^2$. 
Gaps between each the 16 ($\sim 30$ cm)$^2$ modules reduce the usable surface.
Furthermore, most of the 30 MeV to 300 GeV gamma-ray photons that the LAT records trigger the silicon tracker, 
which is $\chi \sim 1$ radiation lengths deep for normal incidence, so the probability that pair production occurs is
somewhat more than $1-1/e = 63\%$. Nevertheless, the effective area of the instrument remains large, of order $10^4$ cm$^2$. 
We take advantage of the combined large area and field of view with a survey strategy: 
we scan the entire sky every two orbits, or eight times per day. See \cite{das_djt_psr_detection} for further
discussion of gamma-ray pulsar detection, and \cite{Pass8} and references therein for details of the LAT. 

\begin{figure}
\includegraphics[width=0.99 \textwidth]{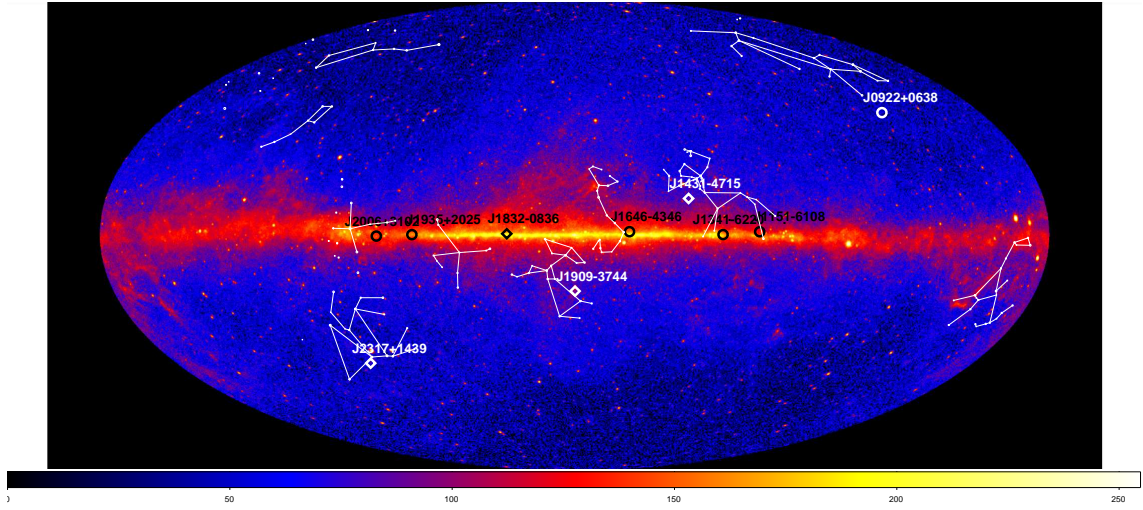}{\centering}
\caption{Eight years of {\em Fermi} all-sky survey data, reprocessed using Pass 8, in an Aitoff-Hammer projection onto Galactic coordinates. 
Only gamma-ray photons with $>1$ GeV are shown. The six new gamma-ray pulsars (circles) and millisecond pulsars (diamonds) are shown, with some visual constellations to guide the eye.
}
\label{skyview}
\end{figure}

Figure \ref{skyview} shows 8 years of gamma-ray photons with at least 1 GeV. The LAT's point spread function (PSF) improves with increasing energy, making the image crisper
than if we include lower energy photons.
The event reconstruction in Figure \ref{skyview} used ``Pass 8'', which extracts the best performance that the collaboration hopes to attain from the instrument.
Before Pass 8, the LAT source catalogs such as 3FGL \cite{3FGL} as well as the second {\em Fermi} LAT pulsar catalog, ``2PC'' \cite{2PC} used only data above 100 MeV.
Our improved mastery of the instrument and of the diffuse and point gamma-ray sources are such that 4FGL and 3PC, both currently in preparation, will lower the minimum
energy to 50 MeV.

The brightest feature in Figure \ref{skyview} is the plane of the Milky Way itself, due to gamma-rays created by cosmic rays incident on interstellar gas and dust. 
Off the plane the point sources are mainly blazars, although 5 to 10\% of them are a sprinkling of millisecond pulsars (MSPs). 
The point sources within the plane are mainly young and recycled pulsars. 
The large number of MSPs that are bright in gamma-rays is one of {\em Fermi}'s most significant discoveries -- roughly half of 
the over two hundred gamma-ray pulsars detected by {\em Fermi} are 
MSPs\footnote{https://confluence.slac.stanford.edu/display/GLAMCOG/Public+List+of+LAT-Detected+Gamma-Ray+Pulsars .}, unexpected by most authors before launch,
and with many interesting consequences.

MSP detection and characterization require accurate and stable timing. 
Section 9 of \cite{OnOrbit} describes ground tests that showed LAT timestamps to be good to at least $\delta t_i = 300$ ns relative to UTC. 
The tests intially showed serious problems which would have completely skewed LAT's pulsar studies, especially for MSPs, as documented on pages 77-78 of \cite{LucasThesis}.
Accurate determination of pulsar rotation phases further requires knowing where the satellite was when the gamma-ray arrived, to an accuracy of roughly $c\delta t_i$, in
three dimensions. {\em Fermi}'s Guidance and Navigation Control system (GNC) does indeed provide the required accuracy, as determined using on-orbit telemetry data
and ground radar tracking of the satellite.

\begin{figure}
\includegraphics[width=0.75 \textwidth, angle=270]{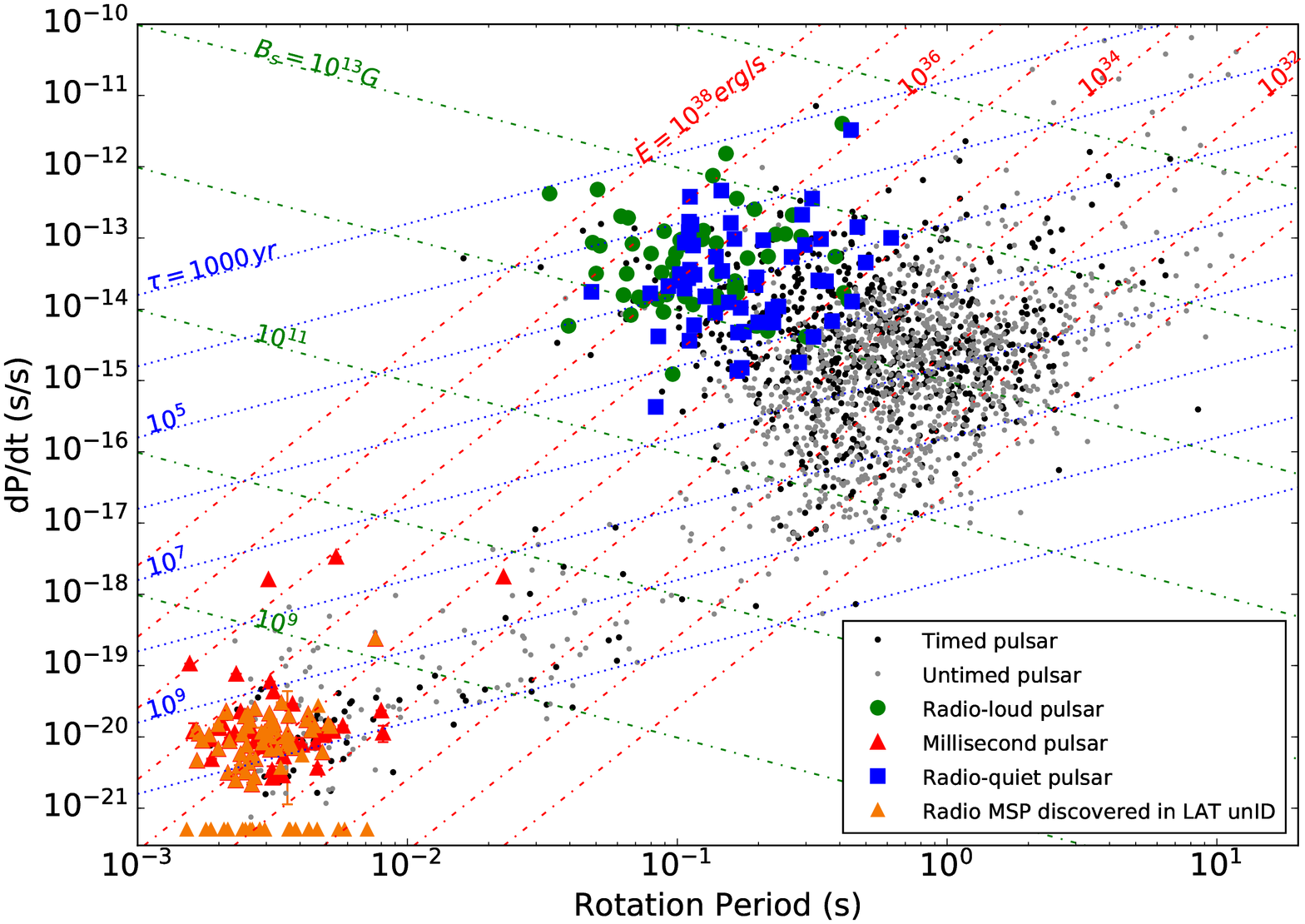}
\caption{P-Pdot diagram with red triangles representing the MSPs, blue squares for radio-quiet pulsars, and green dots for radio-loud young ones. 
The black dots are timed but undetected pulsars in gamma-rays, and grey ones are pulsars for which we have not folded the gamma-rays. 
The horizontal line of triangles indicates the spin period of MSPs found in radio searches of {\em Fermi} unidentified sources,
for which the spindown rate has not yet been published, and gamma-ray pulsations have not yet been seen. 
}
\label{ppdot}
\end{figure}

\section{Discovering Gamma-ray Pulsars}
The spindown versus spin period plot in Figure \ref{ppdot} (updated from 2PC) summarizes the main features of {\em Fermi}'s pulsar harvest. 
Lines of constant spindown power 
$\dot E = 4\pi^2 I_{0} \dot P / P^3$, using $I_{0} = 10^{45}$ g cm$^2$ for the neutron star moment of inertia,
show that to date, all known gamma-ray pulsars have $\dot E > 10^{33}$ erg cm$^{-1}$. 
For non-recycled pulsars (here, the gamma-ray pulsars with period $P > 30$ ms),
this amounts to saying that they are young, with characteristic age $\tau_c = P/2\dot{P}< 30$ Myr, assuming magnetic dipole braking as the only energy-loss mechanism 
and an initial spin period much shorter than the current period. More than half of the non-recycled pulsars have $\tau_c < 100$ kyr.
For the MSPs, \cite{POND} recently confirmed that the minimum $\dot E$ ``deathline'' also still holds for the {\em Fermi} sample: 
observed MSP spindown rates $\dot P$ are subject to a Doppler-like correction that depends on the MSP's distance and proper motion \cite{Shklovskii}.
With better distance estimates, the MSPs that appeared below the deathline fell in line with the others.

Based on the EGRET results \cite{Thompson99}, before launch we anticipated that young, radio-loud pulsars with high $\dot E$ were the ``surest bet'' for
gamma-ray pulsar discovery. Accordingly, we formed a ``Pulsar timing Consortium'' with radio and X-ray astronomers to obtain rotation ephemerides for as many of the known
$\dot E > 10^{34}$ erg cm$^{-1}$ pulsars as possible \cite{TimingForFermi}. The green dots and red triangles in Figure \ref{ppdot} correspond to pulsars for which we used the
known pulsar position to translate the photon arrival times at the satellite to the solar system barycenter, and then used the ephemeris to calculate
a rotational phase for each gamma-ray. When accumulated in a histogram, pulsations may appear, as is in Figures \ref{PSRs} and \ref{MSPs}. 
This method incurs no trials penalities due to searches over $P, \dot P$, position, or other parameters. 
The first new pulsar seen with the LAT, PSR J2021+3651, applied this method \cite{PSR2021LAT} although the discovery was made by AGILE shortly before us \cite{Halpern2008}.
Phase-folding gamma-rays from the direction of known pulsars provided LAT with the discovery of a population of gamma-ray MSPs shortly thereafter \cite{MSP}. 

The blue squares in Figure \ref{ppdot} show pulsars discovered in LAT data, through ``blind'' searches of $P, \dot P$, and position.
Only six are not radio-quiet, i.e., their radio flux density at 1400 MHz is $S_{1400} > 30 \mu$Jy.
The two recent radio-loud blind search pulsars are PSRs J0002+6216 and J0631+0646 \cite{ClarkEatHomeI}.

The position was poorly known in the early days of LAT, unless a well-localized X-ray or radio source could be argued to correspond to the pulsar candidate. 
Poor localization imposes a scan over a grid of positions to ensure that wrong barycentering does not blur the result. The faster the spin period,
the more accurate the position must be, incurring even more trials penalities in blind searches of MSPs. 
Our first discovery, of PSR J0007+7303 in the supernova remnant CTA 1 \cite{CTA1} was quickly followed by many more \cite{BSP}. 

Simple tools sufficed to pick the low-hanging fruit. As the mission continues, pulsars become trickier to discover: 
\cite{SixWeak} discusses the many ways that a pulsar may appear ``weak''. 
{\em Weighting} each photon for its energy-dependent probability of
coming from the (putative) pulsar direction is an especially powerful innovation: previously we suffered trials penalities 
by choosing the cuts on minimum photon energy and maximum angular separation from the nominal pulsar position that maximized 
pulsed significance, generally determined using the H-test statistic applied to the phase histogram \cite{DeJager2010}.
Now, we combine the spectral shape of the candidate source and the background sources with the LAT's PSF to calculate each photon's weight,
and modify the H-test accordingly \cite{KerrWeighted}. 
The {\tt gtsrcprob} routine in the {\em Fermi} Science Tools suite\footnote{https://fermi.gsfc.nasa.gov/ssc/data/} implements weighting.
A simpler tool called {\tt SearchPulsation}, used within the LAT team, applies a few approximations to obtain most of {\tt gtsrcprob}'s power 
without knowing the source spectra.

Innovations also greatly improved the pulsar blind searches in LAT data. 
One is that the sophisticated algorithms developed to search for gravitational wave signals in e.g. LIGO data were
applied to the pulsar searches. Another is that the immense computing power available from citizens' idle PC's was brought to bear by the
Einstein@ Home volunteer computing system: over 10 kyr of CPU time have been used. These and other improvements have doubled the number of known radio-quiet gamma-ray pulsars in the last
few years, as described in \cite{ClarkEatHomeI} and references therein. 
Spectacularly, they have even begun to detect MSPs in blind searches -- the first was PSR J1311-3430 \cite{Pletsch_J1311-3430}.
Those authors are now writing up {\em four} more MSP discoveries.

Figure \ref{ppdot} includes another large discovery set:
MSPs found in radio searches (orange triangles) of the same pulsar-like unidentified gamma-ray source positions as for the blind gamma searches. 
Not only did {\em Fermi} trigger the discovery of over 80 new MSPs in this manner (30\% of all known MSPs outside of globular clusters \footnote{http://astro.phys.wvu.edu/GalacticMSPs/}),
but the {\em Fermi} MSPs are different from those found in radio surveys. 
They are faster, on average, and include large numbers of black widow and redback ``spiders'', that is, accreting and transitional systems \cite{SpiderTransitions}. 
Interstellar scintillation of the fast pulses makes them invisible in a large fraction of radio observations, as do eclipses caused by the inflated companion star
and its violent wind. Consequently, radio surveys would not have found these objects, an illustration of how the {\em Fermi} sky-survey
selects populations complimentary to those found in other searches. See \cite{Deneva_J1048} for a spectacular example of eclipses in PSR J1048+2339. 
In another study, four of six MSPs discovered in {\em Fermi} sources show eclipses \cite{Cromartie_6FermiMSPs}.

After radio periodicity has been discovered, a year or two may lapse before a phase-connected timing solution for the new MSP can be made.
An imprecise MSP position causes annual advances and delays in the pulse arrival times, easily confused with the effects of imprecise $P, \dot P$
and binary parameters. Scintillation and eclipses can further slow a timing campaign, so much that even $\dot P$ can remain mysterious for some time.
The horizontal line of triangles at the lower left of Figure \ref{ppdot} shows pulsars for which we await a $\dot P$  measurement,
as well as an ephemeris adequate for gamma-ray phase folding. We expect nearly all of the MSPs found in searches of gamma-ray
point sources to ultimately reveal gamma pulsations, but it may take years. Currently, about 70\% have.

Finally, the black and gray dots in Figure \ref{ppdot} also merit mention. 
They correspond to known pulsars, listed for the most part in the ATNF database\footnote{http://www.atnf.csiro.au/research/pulsar/psrcat/expert.html} \cite{ATNFcatalog}.
Black means that our radio and X-ray colleagues have provided us with a rotation ephemeris, 
and we gamma phase-folded, but that no significant (H-test $>5\sigma$) pulsations were found. Gray means we have no ephemeris.
Early in the mission we realized that the pre-launch choice of $\dot E > 10^{34}$ erg cm$^{-1}$ was a bit high.
Our radio colleagues provided ephemerides sampling the entire $P, \dot P$ plane, allowing us to plot the fraction of pulsars detected in gamma-rays as a function of $\dot E$
(see Figure 3 in \cite{LaffonNewPSRs} for all pulsars, or Figure 3 in \cite{POND} for MSPs only). 
It rises from zero for $\dot E < 3\times 10^{33}$ erg cm$^{-1}$ to a plateau near 60\% for $\dot E \gtrsim 10^{36}$ erg cm$^{-1}$.
That it remains well below 100\% is presumably because gamma rays are not beamed from all neutron star latitudes.

In the upper right of Figure \ref{ppdot} lie the pulsars with the largest surface magnetic field strengths, $B_{\rm S} = (1.5 I_{0} c^{3} P \dot{P})^{1/2}/2\pi R_{\rm NS}^{3} $ 
(assuming a perfect dipole orthogonal to the rotation axis, and $R_{\rm NS} = 10$ km). Two high $\dot E$, high $B_{\rm S}$ pulsars are seen in gamma rays. 
One is PSR J1119-6127, which recently flared like a magnetar \cite{J1119_magnetarFlare}\footnote{See also E. {G{\"o}{\u g}{\"u}{\c s}}, these proceedings.}.
The other, PSR J1208-6238, was discovered in a gamma-ray blind search and has the 2nd highest known surface B-field \cite{Clark_J1208}.
Overall, however, {\em Fermi} has demonstrated
that magnetars are {\em not} GeV gamma-ray emitters \cite{FermiMagnetarsBis}. 

High $B_{\rm S}$ pulsars do, however, appear to emit brightly in the MeV range \cite{KuiperHermsen2015}.
Recently, one of them, PSR J1846-0258 was detected below 100 MeV in the LAT data \cite{Kuiper_J1846-0258}\footnote{See also L. Kuiper, these proceedings.}.
Of 18 pulsars seen in the range 20 keV $< E_\gamma <$ 30 MeV, only half are also seen in {\em Fermi} LAT due in part to the spectral peak in the MeV range. 
All the $E_\gamma <100$ MeV rotation powered pulsars have $\dot E>4\times 10^{36}$ erg/s.

High $B_{\rm S}$ induces braking so strong (large $\dot P$) that $\ddot P$ may also be measurable (timing noise generally dominates $\ddot P$).
Indeed, PSR J1208-6238 mentioned above has yielded an accurate braking index, $n = 2 - P\ddot P / \dot P^2 = 2.598$.
{\em Fermi} LAT detected gamma-ray pulsations from PSR B0540-69 in the Large Magellanic Cloud (LMC) thanks to RXTE X-ray timing \cite{B0540_gamma_discovery}.
We then began a timing campaign with {\em Swift} that revealed that $\dot E$ had increased by a spectacular 36\% \cite{B0540_spindown_change},
and then that $n$ had changed from its historical value of $n = 2.129 \pm 0.012$ to a stunningly low value
of $n = 0.031 \pm 0.013$ \cite{B0540_braking_index}. An interpretation is that the neutron star's B-field is `unburying' itself. 
Since the jump in late 2011, the B-field would be emerging more quickly,
with a 831 year time scale as compared to the 3859 year rate seen before.
An interesting recent discussion of braking indices is provided by \cite{SimonArisBraking}. 
{\em Swift} continues to time PSR B0540-69 and LAT will soon be able to see whether the gamma-ray profile also changed after the state shift.

\section{Recent Discoveries}
\begin{figure}
\includegraphics[width=1.1 \textwidth]{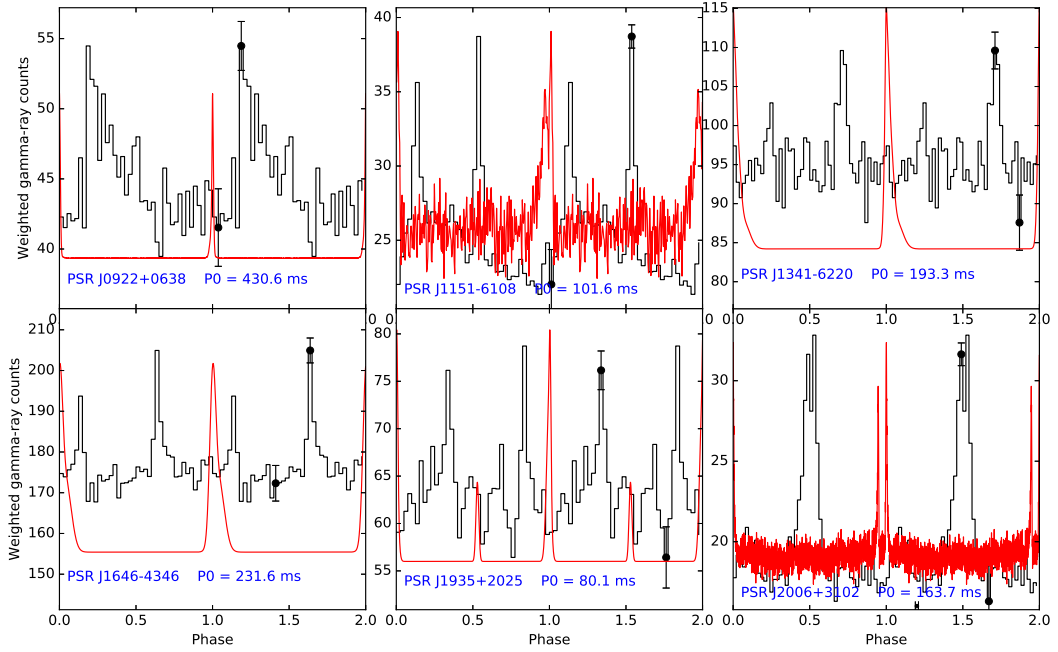}{\centering}
\caption{Six recently discovered young gamma-ray pulsars. The phase interval is shown twice, for clarity. 
The largest and smallest error bars are shown for each gamma-ray phase histogram. The gamma-ray histograms are phase-aligned with the $1.4$ GHz radio profiles; overlaid in red.
}
\label{PSRs}
\end{figure}

\begin{figure}
\includegraphics[width=0.8 \textwidth]{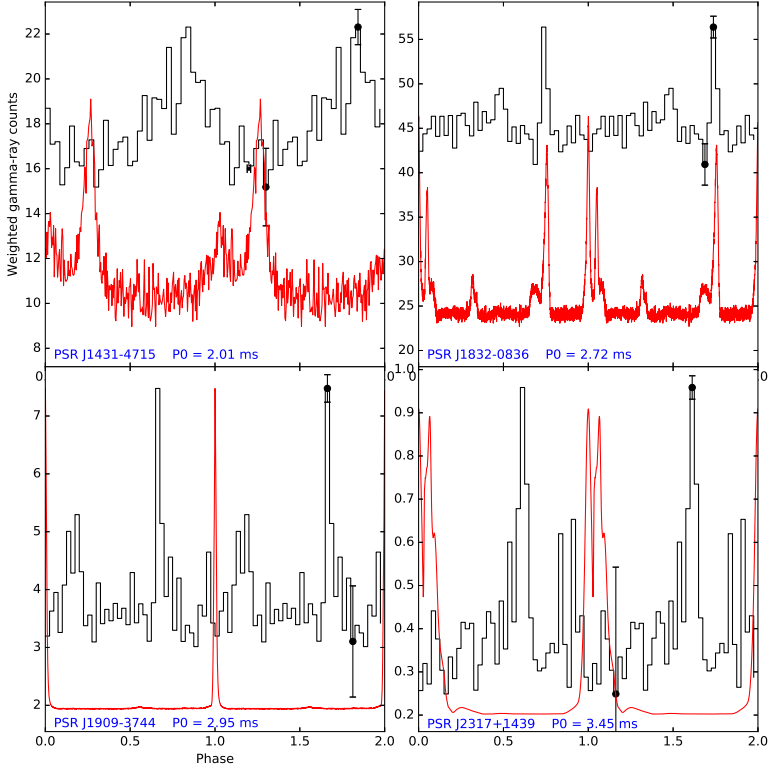}{\centering}
\caption{Four recently discovered gamma-ray millisecond pulsars. Details as in the previous figure.}
\label{MSPs}
\end{figure}

Table \ref{tbl-charPSR} lists 10 pulsars for which gamma-ray pulsations were discovered by phase-folding LAT photons using rotation
ephemerides obtained from timing at radio telescopes.
The references to the radio discovery papers are the ATNF mnemonics, given explicitly for each pulsar, below.
We used Pass 8 data through MJD 57850, $E_\gamma > 100$ MeV, within $2^\circ$ of the pulsar's radio timing position, 
except for PSR J0922+0638 for which we extended the data region to $5^\circ$.
Some of the ephemerides do not cover the entire mission epoch, but gamma-ray pulse phase may remain steady outside of the radio validity period.
The table lists how much data was folded. 
Figure \ref{PSRs} shows the six young gamma-ray pulsars, and
Figure \ref{MSPs} shows the four gamma-ray millisecond pulsars. 
Phase 0 is set to the peak of the radio signal, with gamma-ray phases calculated using the {\em tempo2} \cite{Hobbs2006} {\tt fermi}
plugin\footnote{http://fermi.gsfc.nasa.gov/ssc/data/analysis/user/Fermi\_plug\_doc.pdf}, with the same ephemeris as for the radio profile.

The first distance, $d$, is derived from the pulsar's dispersion measure (DM) using the ymw16 model \cite{ymw16} and the second uses NE2001 \cite{Cordes2002} as default, 
possibly over-ruled by a ``better'' distance obtained from the literature, as judged by the ATNF database team.
We examined the pulsars' luminosity $ L_\gamma = 4\pi d^2 f_\Omega G_{100}$ using preliminary measurements of the integral energy flux above 100 MeV, $G_{100}$,
obtained from a working version of the 4FGL catalog being prepared by the LAT team. 
The efficiencies all have typical low values, $1\% < \eta= L_\gamma/\dot E < 25\%$. This suggests that PSR J1341-6220, with $ \eta \simeq 7\%$ is indeed plausibly at such a large distance.
We set the gamma-ray beaming fraction to $f_\Omega = 1$, as defined in 2PC.
3PC will provide $L_\gamma$ for all gamma-ray pulsars with known distances (excluding, alas, most radio-quiet pulsars), as well as careful determinations of the peak positions and widths.
In the following we discuss each pulsar.

\subsection{Young pulsars}

{\bf PSR J0922+0638:}  mlt+78  \cite{Molonglo}.     
One of the lowest $\dot E$ young radio-loud gamma-ray pulsars so far.
The slowest young radio-loud gamma-ray pulsar so far (right-most green dots in Figure \ref{ppdot}), although five radio-quiet pulsars are slower. 
Well off the plane, it appeared as a steady source with a low Test Statistic (TS) of 19 in an early version of the LAT 7 year catalog, inciting us to look closely at its phase folding.
Due to its faintness, we needed to extend the data sample to $5^\circ$, possible because of its high latitude and the absence of nearby sources.

{\bf PSR J1151-6108}  ncb+15 \cite{Ng2015_HTRUN}. 
This faint radio pulsar has a high TS in spite of being in a high-background region, which allowed a spectral analysis showing a cut-off at a few GeV, typical of pulsars.
Discovered in mid-2013, only $2.2$ years of radio timing data were available at first, but gamma-ray folding revealed pulsations. 
Parkes resumed observations, allowing folding to the present. 
The pulsar is too faint in gamma-rays to be able to generate times-of-arrival by the methods of \cite{LATtiming}.
However, maximizing H-test while scanning a $P, \dot P$ range near the radio ephemeris values allowed us to extend the signal to the beginning of the mission.
 
We also phase-folded PSRs J1616-5017 and J1731-3325, $\dot E>10^{34}$ erg/s pulsars discovered in the survey reported in \cite{Ng2015_HTRUN}. No pulsations appeared.

{\bf PSR J1341-6220} mdt85  \cite{ManchesterMSPs_1985}.
$13^\circ$ closer to the Galactic center and even closer to the plane than J1151-6108, and apparently quite distant, the pulsar is detectable because its high $\dot E$ leads to a large luminosity $L_\gamma$
but also because its peaks are narrow.
The gamma-ray source $0.1^\circ$ from the pulsar had been thought to be the supernova remnant SNR G$308.8-0.1$ \cite{KaspiJ1341-6220_1992}. 
However, LAT does not detect this SNR \cite{SNRcat} and the pulsar dominates the gamma emission.

{\bf PSR J1646-4346} jlm+92 \cite{JohnstonSouthernSurvey1992}.
Closer still to the Galactic center and again very near the plane,
a clear pulsed signal appeared for this pulsar (as for PSR J1341-6220) when we folded 300 pulsars using Parkes ephemerides updated after a hiatus of about two years, intended to let
gamma-ray statistics accumulate.  Again, the narrowness of the peaks helps overcome the high background.

{\bf PSR J1935+2025}  lcm13 \cite{LorimerParkesTiming_2013}.
A rare radio interpulse pulsar \cite{interpulses} (they don't cite this pulsar because it was discovered later), with high $\dot E$, straggling the plane, and fairly distant. 
The pulsar lies at the edge of the error ellipse of a weak gamma-ray source that has, within the poor statistics, a possibly pulsar-like spectrum.
Radio interpulses place tight constraints on the inclination of the neutron star's rotation and magnetic axes, as well as the inclination to the observer's line of sight.
Combined with the constraints imposed by the gamma-ray pulse profile \cite{AtlasII}, their particular interest is to allow stringent tests of magnetospheric models.
Timing observations for this pulsar ended in 2011 and the pulsed gamma-ray signal can not be seen beyond May 2011 (MJD 55700). It is too faint to time with the LAT data.

{\bf PSR J2006+3102} nab+13 \cite{nab+13} NRT.
Another pulsar discovered after launch, after the initial list of LAT targets had been defined \cite{TimingForFermi}, it is by far the brightest of the gamma-ray pulsars presented here.

\subsection{Millisecond pulsars}

{\bf PSR J1431-4715} btb+15 \cite{BatesHTRUN-XI_2015}.
Timing for this ``redback'' binary began in July 2011, after confirmation of the discovery observations. 
The ephemeris obtained by avoiding the eclipse epochs is stable enough that gamma-ray pulsations are clear beginning two years earlier.
The suggestion in \cite{BatesHTRUN-XI_2015} that the pulsar could be detectable with the {\em Fermi} LAT was correct.

{\bf PSR J1832-0836} bbb+13 \cite{BurgayHTRUN-VII_2013}.
The narrow gamma-ray peak, FWHM $< 30 \mu$s, rivals that of the first ever black widow pulsar discovered, PSR B1957+20 \cite{GuillemotBlackWidow_2012}, again presumably due to
caustic emission. It is aligned with a similarly narrow radio peak.
PSR J1832-0836 is the only MSP presented here that lies right on the plane, near the Galactic center. The narrow peak surely helped making this faint source detectable
despite the intense background, an element that must be considered when searching for a putative population of MSPs in the Galactic bulge \cite{LATbulgeMSPs_2017}.

{\bf PSR J1909-3744}  jbv+03 \cite{JacobyJ1909-3744_2003}.
A stable pulsar with a sharp, narrow radio peak, it is included in the Pulsar Timing Array searches for gravitational waves \cite{EPTA}.
The nearest 7-year source is $0.81^\circ$ away with TS$=45$ and $0.2^\circ$ uncertainty,
we thus have another example of a pulsar for which only pulsed emission is detected.

{\bf PSR J2317+1439} cnt93  \cite{cnt93}.
The blazar 3C $454.3$ nearly made us miss this pulsar (the blazar is visible in Figure \ref{skyview} under the `+' of the pulsar name). 
The pulsar was associated with a TS$=27$ source beginning with a 6 year LAT preliminary source list, prompting our close attention.
The pulsed signal was weak and, suspiciously, H-test does not increase uniformly in time as it generally does for LAT pulsars. 
It turns out to be less than $5^\circ$ from the blazar, which has had phenomenal flares throughout the {\em Fermi} mission \cite{FlaringBlazar}. 
Even at that angular distance from the blazar, 
the flares increased the background for the pulsar enough to decrease the detection rate at times.

$\dot E$ in Table \ref{tbl-charPSR} is uncorrected for the Shklovskii effect.
The corrected value provided by ATNF is $\dot E_I = 2.2 \times 10^{33}$ erg/s, slightly lower than those reported in \cite{POND},
suggesting a slightly lower gamma-ray deathline than previously known.

\begin{table*}[ht]
\caption{Six young (top) and four recycled (bottom) radio pulsars for which we have discovered gamma-ray pulsations.
Distances are the ATNF ``DIST'' and ``DIST1'' variables (see text). See text for the discovery references. The rotation ephemeris used in this
work was obtained from radio timing with: NRT (Nan\c cay Radio Telescope) \cite{Cognard2011} ; 
PKS (Parkes Radio Telescope) \cite{ParkesFermiTiming}, extended in some cases to longer durations using LAT gamma-ray timing \cite{LATtiming}.
\label{tbl-charPSR}}
\begin{scriptsize}
\begin{tabular}{lrrrrrrrrrr}
PSR &        $l$       &  $b$     & $P$ & Distance  &$\dot E /10^{34}$&$S_{1400}$&H-test     &Years &Radio    & Timing \\
    &     ($^\circ$)   &($^\circ$)&(ms) & (kpc)     & (erg s$^{-1}$)  & (mJy)	 & signif.   &folded&discovery& \\
 J0922+0638   & 225.42 & 36.39    &430.6& 1.1 / 1.1 & 0.66	      &  4.2	 &$6.3\sigma$& 8.7  &mlt+78   & NRT\\ 
 J1151-6108 & 295.81 & 0.91     &101.6& 2.2 / 4.2 & 38.65	      & 0.06	 &$7.0\sigma$& 8.7  &ncb+15   & PKS\\ 
 J1341-6220 & 308.73 & -0.03    &193.3& 12.6 /11.1& 138.01	      & 1.9      &$5.9\sigma$& 7.2  &mdt85    & PKS\\ 
 J1646-4346 & 341.11 & 0.97     &231.7& 6.2 / 5.8 & 35.70	      & 0.98	 &$>8\sigma$ & 7.2  &jlm+92   & PKS\\ 
 J1935+2025   & 56.05  & -0.05    &80.0 & 4.6 / 6.2 & 463.89	      & 0.527    &$7.3\sigma$& 2.8  &lcm13    & PKS\\ 
 J2006+3102   & 68.67  & -0.53    &163.7& 6.0 / 4.7 & 22.38	      & 0.27     &$>8\sigma$ & 7.4  &nab+13   & NRT\\ 
 & & & & & &  & & & & \\
 J1431-4715 & 320.05 & 12.25    &2.01 &  1.8 / 1.6& 6.45	      & 0.73     &$6.2\sigma$& 7.5  &btb+15   & PKS\\ 
 J1832-0836 & 23.11  & 0.26     &2.72 &  0.8 / 1.1& 1.69	      & 1.10	 &$6.3\sigma$& 7.3  &bbb+13   & NRT\\ 
 J1909-3744 & 359.73 & -19.60   &2.95 &  1.1 / 1.1& 0.43	      & 2.10     &$7.2\sigma$& 5.7  &jbv+03   & NRT\\ 
 J2317+1439   & 91.36  & -42.36   &3.45 &  1.4 / 1.4& 0.28	      & 4.00     &$5.1\sigma$& 7.7  &cnt93    & NRT\\ 
\end{tabular}
\end{scriptsize}
\end{table*}

\section{Prospects}
Ending its $9^{th}$ year in orbit, the {\em Fermi} LAT continues to discover about 25 gamma-ray pulsars per year.
Different streams feed this flow: radio and gamma-ray pulsation searches at the positions of unidentified LAT sources are two.
Here, we presented recent results from a third stream: using radio ephemerides to phase-fold the gamma-rays from as many
pulsars as possible. Pulsations can appear even if the gamma ray source, integrated over phase, is faint or undetected.
All but two or three of the new pulsars presented here have very narrow gamma-ray pulses, helping discovery.
In \cite{SixWeak} however we showed faint pulsars found in spite of very {\em broad} pulses.

New methods appear and trigger new bursts of discoveries. A current example is VLA searches for
steep spectrum radio sources spatially co-located with LAT UnId's -- pulsar candidates are not only identified, but also superbly localized,
easing  periodicity searches \cite{FrailUnIds,FrailGeVExcess}.

The varied searches yield varied pulsar types. Selection biases vary for the various searches. 
We can hope that the resulting gamma-ray sample thus represents the true population as well as possible, important when trying, for example,
to extrapolate to the population of unresolved pulsars that might mimic a signal from supersymmetric dark matter \cite{LATbulgeMSPs_2017}.
Comparing the variety of profile shapes and efficiencies $\eta$ with emission model predictions, as in \cite{AtlasII}, also requires that
the biases in the observed sample be understood.

Pass 8 \cite{Pass8}, guided by INTEGRAL observations, is allowing a $4^{th}$ stream to develop: the search for gamma-ray pulsars with spectral cut-offs
below a few hundred MeV. PSR J1846-0258 may be a harbinger of yet another gamma-ray pulsar sub-class, straddling the LAT and INTEGRAL energy ranges \cite{Kuiper_J1846-0258}.
The launch of a new instrument with good sensitivity and angular resolution in the MeV range would surely result in yet more gamma-ray pulsar discoveries.
Possibilities are HARPO \cite{Harpo} and e-ASTROGAM \cite{eAstrogam}.

3PC, the $3^{rd}$ {\em Fermi} LAT catalog of gamma-ray pulsars, will detail over 215 objects. It follows 2PC \cite{2PC},
and will rely heavily on the results in 4FGL, the 8-year {\em Fermi} LAT source catalog, successor to 3FGL \cite{3FGL}. 
Two thorough reviews for interested readers are \cite{CaraveoRevolution, GrenierHardingGold}.

{\bf Acknowledgements:} 
The Nan\c{c}ay Radio Observatory is operated by the Paris Observatory, associated with the French Centre National de la Recherche Scientifique (CNRS).
The Parkes radio telescope is part of the Australia Telescope which is funded by the Commonwealth Government for operation as a National Facility managed by CSIRO. 
Fermi work at NRL is supported by NASA.

The \textit{Fermi}-LAT Collaboration acknowledges support for LAT development, operation and data analysis from NASA and DOE (United States), CEA/Irfu and IN2P3/CNRS (France), 
ASI and INFN (Italy), MEXT, KEK, and JAXA (Japan), and the K.A.~Wallenberg Foundation, the Swedish Research Council and the National Space Board (Sweden). 
Science analysis support in the operations phase from INAF (Italy) and CNES (France) is also gratefully acknowledged. 
This work performed in part under DOE Contract DE-AC02-76SF00515.

\bibliographystyle{Science}

\bibliography{2ndPulsarCatalog}


\end{document}